\documentclass[pre,aps,twocolumn,showpacs]{revtex4}
\usepackage{epsfig}
\usepackage{epsf}
\begin{document}
\newcommand{\D}{\displaystyle}

\pre{{\bf 55}, 6186-6196 (1997)}

\title{Conformal Mapping on Rough Boundaries \\II: 
Applications to bi-harmonic problems}
\author{Damien Vandembroucq$^{(1,2)}$ and St\'ephane Roux$^{(2)}$ }
\altaffiliation[present address ]{Unit\'e Mixte CNRS/Saint-Gobain, 39 Quai Lucien Lefranc 93303 Aubervilliers FRANCE}
\email{damien.vdb@saint-gobain.com, stephane.roux@saint-gobain.com}
\affiliation{{\small (1)} Department of Applied Mathematics and
  Theoretical Physics\\
Cambridge CB3 9EW, UK \\
{\small (2)} Laboratoire de Physique et M\'ecanique des 
Milieux H\'et\'erog\`enes,\\
Ecole Sup\'erieure de Physique et de 
Chimie Industrielles,\\
10 rue Vauquelin, 75231 Paris Cedex 05, France}

\begin{abstract}
We use a conformal mapping method introduced in a companion paper
\cite{VdbR96a} to study the properties of bi-harmonic
fields in the vicinity of rough boundaries. We focus our analysis on
two different situations where such bi-harmonic problems are
encountered: a Stokes flow near a rough wall and the stress
distribution on the rough interface of a material in uni-axial
tension. We perform a complete numerical solution of these
two-dimensional problems for any univalued rough surfaces. We present
results for sinusoidal and self-affine surface whose slope can locally
reach 2.5. Beyond the numerical solution we present perturbative
solutions of these problems. We show in particular that at first order
in roughness amplitude, the surface stress of a material in uni-axial
tension can be directly obtained from the Hilbert transform of the
local slope. In case of self-affine surfaces, we show that the stress
distribution presents, for large stresses, a power law tail whose
exponent continuously depends on the roughness amplitude.  
\end{abstract}


\maketitle

\section{Introduction}

In a companion paper\cite{VdbR96a}, we have presented a conformal
mapping technique that allows us to map any 2D medium bounded by a
rough boundary onto a half-plane. This method is based on the
iterative use of FFT transforms and is extremely fast and efficient
provided that the local slope of the interface remains lower than one.
When the maximum slope exceeds one this algorithm, similar in spirit
to a direct iteration technique well suited to circular
geometries\cite{Gut81,Henrici3}, can no longer be used in its original
form.  Underrelaxation\cite{Gut83} suffices however to make it
convergent for boundaries having large slopes.  Beyond the
determination of a conformal mapping for a given rough interface we
have also shown in Ref.\cite{VdbR96a} how to generate directly
mappings onto self-affine rough interfaces of chosen roughness
exponent. Self-affine formalism is an anisotropic scaling invariance
known to give a good description of real surfaces such as fracture
surfaces\cite{Mandelbrot84Nat,EBouchaud90EPL,Malloy92PRL}. This
statistical property of fracture surfaces is of great interest in the
study of friction or transport processes in geological
faults\cite{Schmittbuhlphd,Plouphd}.

Building a conformal mapping is entirely equivalent to solving a
harmonic problem with a uniform potential (or field) condition on the
boundary. We used extensively this property in Ref.\cite{VdbR96a} to
study stationary heat flows in the vicinity of a rough boundary and we
focussed on the case of self-affine surfaces where we were able to
compute the exact correlation between local surface field and height
profile. We gave also a special emphasis to the problem of the
location of the plane interface equivalent to the rough one at
infinity. It turned out that the conformal mapping technique provides
a very direct means of computing the shift between the plane
equivalent interface and the mean plane of the rough interface.

The range of applications of this first study naturally covers fields
where Laplace equation appears: electrostatics, concentration
diffusion, antiplane elasticity... In this second paper, we are
specifically concerned with the case of bi-harmonic problems.  The
method we propose leads to the solution of the bi-harmonic field
through the inversion of a linear system, as most other alternative
numerical approaches (e.g. boundary elements method, spectral
method...) but in contrast to the latters, the linear system to invert
is naturally well conditionned and of rather modest size (N equations
for N Fourier modes in the conformal mapping method) in contrast with
direct spectral methods (4N unknowns).

Moreover, following the first step of our algorithm analytically
allows to obtain systematic perturbation expansion results.

After recalling our main results about conformal mapping in the first
section, we deal successively with two important examples of
bi-harmonic problems: Stokes flows and plane elasticity. The second
section is thus devoted to the study of a stationary Stokes ({\it
  i.e.} low Reynolds number) flow close to a rough boundary. We shall
also develop in this section the paradigm of the equivalent
``no-slip'' plane interface. In the third section we point out the
problem of the stress distribution in a two-dimensional material
bounded by a rough boundary and submitted to a uni-axial tension. The
study of such situations is of particular interest for computing the
influence of the roughness on the rupture probability law of brittle
materials. We show that this problem is formally identical to that
previously solved for the Stokes flow. We pay further a special
attention to the case of self-affine surfaces and we present in this
section numerical results that suggest that for large stresses, the
surface stress statistical distribution law presents a power law behavior.

\section{Conformal mapping on rough boundaries}
We recall here the essential results described in Ref. \cite{VdbR96a}.
The aim of this section is to conformally map a half-plane onto a
two-dimensional domain bounded by a rough interface. We first recall
briefly how to build a conformal mapping well-suited to a given rough
interface.  This problem can be written under form similar as that of
a ``Theodorsen problem'' \cite{Henrici3}; in the semi-infinite
geometry we deal with, it can be solved with an iterative algorithm
using fast Fourier transforms \cite{VdbR96a}. The second part of this
section is focussed on the specific case of self-affine interfaces. It
turns out, indeed, that very simple constraints on a conformal mapping
allow to generate directly a two-dimensional domain bounded by a
self-affine interface of chosen roughness exponent. This property is
of particular interest in statistical studies. We could thus establish
in Ref.\cite{VdbR96a} the correlation between the norm of a harmonic
field at a self-affine interface and the height of this latter.

\begin{figure}[tbp]
 \centerline{
    \epsfig{file=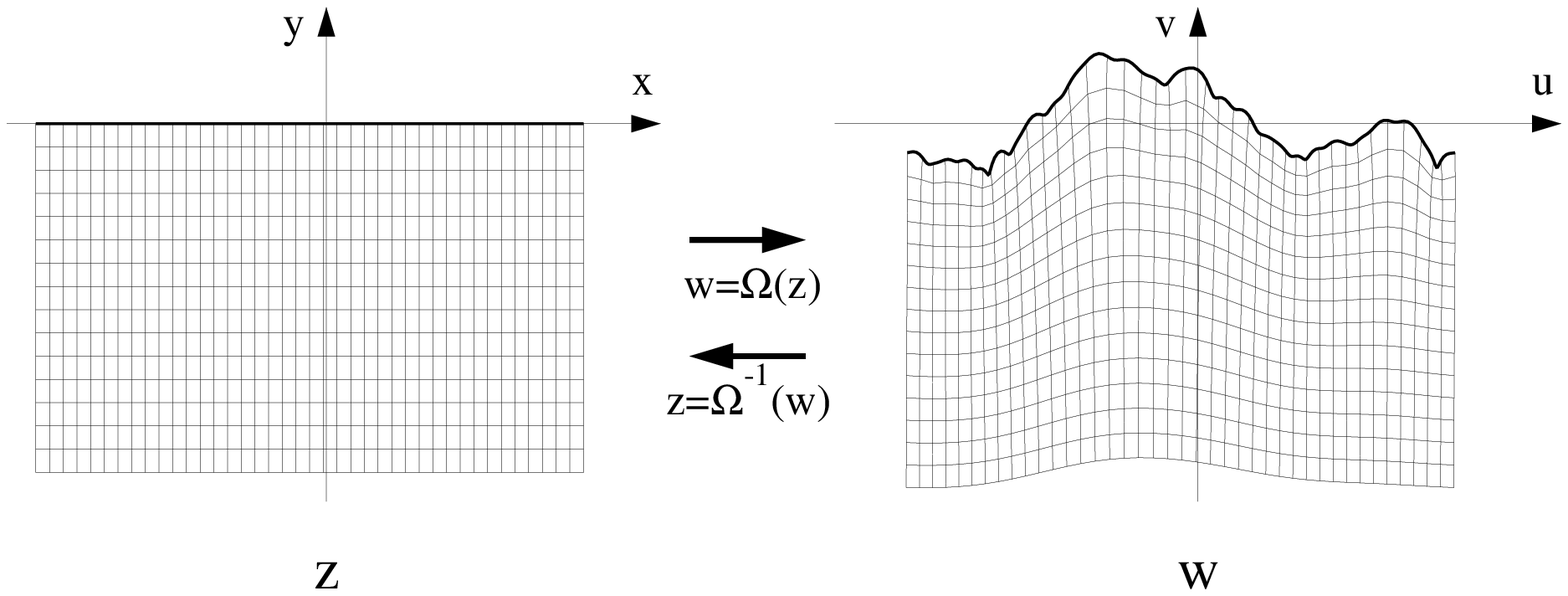, width=0.85\hsize,angle=0}} 
\caption{A schematic illustration of the mapping $\Omega$ which maps the
semi-infinite plane ${\cal D}$ onto the domain limited by a rough
interface ${\cal E}$.}  
\label{fi:method}
\end{figure}

\subsection{Notations}

As illustrated in Fig. \ref{fi:method}, we are seeking for a mapping
from a half-plane onto a two-dimensional domain bounded by a rough
interface. In the following we place our study in the framework of the
complex analysis. We consider then the lower half-plane ${\cal D}$
whose complex coordinate $z=x+iy$ is such that ${\Im m} z \le 0$ and
the two-dimensional domain ${\cal E}$ bounded by the rough interface
$\partial {\cal E}$, we call $w=u+iv$ the complex coordinate in ${\cal
  E}$. We are seeking for a mapping $\Omega$ from ${\cal D}$ onto ${\cal
  E}$. We restrict our study to mappings $\Omega$ that are bijective
holomorphic functions, {\it i.e.} $\Omega$ only depends on the
variable $z$ and not on its conjugate ${\overline z}=x-iy$. The
transformations associated with such functions are said to be
conformal in the sense that they preserve locally the angles. Let us
now take advantage of the semi-infinite geometry we have to deal with.
The two domains we consider are very similar apart from the region
close to the boundary.  Far from this one, $\Omega$ is essentially the
Identity and we can  write:
\begin{equation}
\Omega(z)=z+\omega(z)\;,
\end{equation}
where the perturbation $\omega$ decreases with depth and takes
non negligible 
 values  only in the close vicinity of the interface. In the
following, we consider periodic  interfaces in order to minimize edge-effects. Without loss of
generality, let us choose $2\pi$ to be the lateral period. A natural
form of $\Omega$ is then:
\begin{equation}
\Omega(z)=z+\omega(z)=z+ \sum_{k=0}^\infty \omega_k e^{-2i kz}\;.
\end{equation}
where $\omega$ is expanded on a basis of evanescent modes.

\subsection{Computing the mapping for an imposed interface}

We consider a single valued interface $\partial{\cal E}$. Let $h$ be
the real function giving the interface geometry, for all point
$w=u+iv$ of $\partial{\cal E}$,
\begin{equation}
 v=h(u)\;.
\end{equation}
The mapping function $\Omega$ is such that $\Omega
\left(\partial{\cal D}\right)=\partial{\cal E}$, {\it i.e}
\begin{equation} \label{para}
\begin{array}{ll}
\vspace{4pt}
&{\D h\left(u \right) = {\Im m} \left(\sum_{k=0}^\infty \omega_k
e^{-ikx} \right)}\;,\\
{\rm where} &{\D u=x+{\Re e} \left(\sum_{k=0}^\infty \omega_k
e^{-ikx} \right)}\;.
\end{array}
\end{equation}
The first equation is here very close to a Fourier transform except
that we have $h(u)$ instead of just $h(x)$ in the first term. This
formal proximity can be used to build an iterative algorithm. For
sufficiently small roughness, we can see from the second equation that
$x$ is an approximation of $u$ at zeroth order. A direct Fourier
transform of the profile $h(x)$ allows then a first approximation
$\{\omega^{(0)}_k\}$ of the coefficients $\omega_k$. The latter can be
used to correct the previous approximation of $u(x)$, using
(\ref{para}b) gives then a following estimation of the non-uniform
sampling $u(x)$ and of the coefficients $\omega_k$ via the Fourier
transform of $h[u(x)]$. In appendix A, we give a more formal
presentation of this algorithm. It turns out that this iterative
technique converges provided that the maximum slope of the profile
remains below unity.  The technique can  be
used for any rough single-valued interface.  For profiles whose
maximum slope exceeds one, the algorithm can be made convergent with
slight modifications such that the use of under-relaxation techniques.
We refer the reader to Ref. \cite{VdbR96a} for more details on the
convergence and the stability analysis in this specific framework.
Extensive studies of this technique in case of cicular geometry are
available in Ref. \cite{Gut81,Gut83}.

\subsection{Conformal mapping on self-affine interfaces}
As pointed out above, the algorithm we have just described is suited
to any rough interface. It is in particular possible to build
conformal mappings associated with self-affine interfaces. The latter
are defined by their scaling invariance properties: an interface
described by the equation $y=h(x)$ is said to be self-affine if it remains
(statistically) invariant under the transformations
\begin{equation}
\left\{
\begin{array}{rl}
x & \to\lambda x \;,\\ 
y & \to\lambda^\zeta y \;,
\end{array}
\right.
\end{equation}
for all values of the real parameter $\lambda$. The exponent $\zeta$
is called the ``Hurst'' or roughness exponent. It is characteristic of
the scaling invariance. From this property, we derive easily that
\begin{equation}  \label{eqsa}
\langle (h(x)-h(x+\delta))^2\rangle=C^2\delta^{2\zeta} \;,
\end{equation}
where $C$ is a prefactor. A simple Fourier transform gives then the
power density spectrum of the rough self-affine profile.
\begin{equation}  \label{eqsaps}
P(k) \propto k^{-1-2\zeta}\;.
\end{equation}
When using the algorithm previously described to map a self-affine
interface, the first guess for the mapping function coefficients
$\omega_k$ is thus such that:
\begin{equation}  
{\D \omega_k=2i a_k \propto k^{-\frac{1}{2}-\zeta}} x_k \;,
\end{equation}
where the $a_k$ are the coefficients of the Fourier transform of the
profile and $x_k$ are $k$ independent random variables. It turns out that this power law behaviour is not altered by
the following steps of the algorithm. 
In a symmetric way, we can impose, without any further restriction,
the $\omega_k$ to follow a power law and have a look at the interface
generated. We choose thus
\begin{equation}  
\omega_k=A\epsilon_k k^{-1/2 - \zeta} \;,
\end{equation}
where $\epsilon_k$ are random Gaussian variable with 0 mean and unit
variance for the real and imaginary part independently. We have however
to note that nothing prescribes {\it a priori} that the
function obtained is bijective. From the parametrical expression of the
interface $\partial {\cal E}$, we can write 
\begin{equation}
\mathop{\Re e} \left[ {\frac{\partial \Omega }{\partial x}} (x+i0) \right] =
1 + A. \mathop{\Im m} \left[ \sum_k \epsilon_k k^{1/2 - \zeta} e^{-ikx}
\right] \;,
\end{equation}
and to guarantee that $\partial {\cal E}$ remains single-valued we
 have to
choose amplitudes lower than the threshold value
$A_{max}$ where 
\begin{equation}  \label{eqampmax}
A_{max}={\frac{-1 }{{\mathop{\Im m} \left[ \sum_k \epsilon_k k^{1/2 - \zeta}
e^{-ikx} \right]}}}\;.
\end{equation}
We checked numerically (see Ref. \cite{VdbR96a} that the power spectrum of such
synthetic profiles was indeed a power law with the expected exponent.

\section{Stationary Stokes Flow in the vicinity of a rough wall}

The Stokes equation describes fluid flows at low Reynolds numbers. We
address in this section the problem of a stationary Stokes flow in the
vicinity of a rough boundary. The study of such flows can be of great
technological interest in the case of convective transport processes
\cite{Hig85}: one can think of problems of surface deposition or erosion.
In the same spirit, the occurrence of recirculating eddies can render
very difficult the decontamination of a polluted surface ; contaminants
particles can be captured by diffusion in a cavity and remain trapped
in it for an arbitrary long time. From a more fundamantal point of
view, in experiments consisting of tracing particles passively advected
in a flow, the same process can lead to non gaussian statistics of the
arrival time of tracer particles.

We consider a semi-infinite 2D
geometry with periodic lateral boundary conditions and a unit shear
rate at infinity. Far from being specific, the results obtained in
this context can easily be extended for any case of stationary shear
flow in the framework of a double scale analysis. Considering a shear
flow of shear rate $\dot{\gamma}$ and an interface of typical boundary
$\varepsilon$, where the the velocity and pressure fields ${\bf U}$
and $p$ obey the usual Navier Stokes equation, we can, following
Richardson \cite{Richardson73JFM}, define an inner problem where the
reduced non dimensional variables obey, at first order in $\varepsilon$,
a simple Stokes equation in a semi-infinite geometry. In the following
we present a solution of this Stokes equation that can be rewritten as
a biharmonic equation for the stream function. This solution only
requires the knowledge of a conformal mapping $\Omega $ from the lower
half plane ${\cal D}$ onto the actual space ${\cal E}$ bounded by the
rough interface $\partial {\cal E}$ and the inversion of a well
conditioned linear system ; it can thus be applied to any single-valued
rough interface. We focus this brief study on the problem of the
determination of the location of a plane boundary equivalent to the
rough one at infinity. This problem is equivalent to the one of the
replacement of the no-slip condition on a rough interface by a back
flow condition (to be determined) on the mean plane. We compare our
results with those of Tuck and Kouzoubov \cite{Tuck95} who developped in the {\it
  actual space } a method similar to ours in spirit. Recent results
about Stokes flows near rough boundaries can also be found in
references \cite{Hig85,Jansons88,Poz95}; in most of them 
the Stokes equation is solved using boundary elements methods (see for
instance the review of Pozrikidis \cite{Pozbook}).

\subsection{General solution}
We adress here the problem of a unit shear Stokes flow in the vicinity
of a rough boundary. Let us call $\Psi(w)$ the  stream-function
associated to the veolociy field ${\bf b}$ in the actual space ${\cal
  E}$. We have by definition
\begin{equation}
b_u=\frac{\partial \Psi}{\partial v}\;, \quad b_v=-\frac{\partial
  \Psi}{\partial u} \;.
\end{equation}
In a stream function formalism, the Stokes equation is reduced  to
a simple bi-harmonic equation. Taking into account the boundary
conditions {\it i.e.} no slip on the interface and unit shear rate at
infinity, $\Psi(w)$ has to be solution of the following problem:
\begin{equation}  \label{eqperbi0}
\left\{ 
\begin{array}{rlcc}
\nabla_w^4 \Psi (w)& = 0\quad & {\rm in}\quad & {\cal E} \\ 
{\bf \nabla}_{w} \Psi (w)& ={\bf 0} \quad & {\rm on}\quad & \partial{\cal E} \\ 
\Psi(w) & \sim  \frac{1}{2}v^2 \quad & {\rm as}\quad & v \to - \infty
\end{array}
\right.
\end{equation}
The essential difficulty obviously lies in the no slip condition on
the interface ; the use of a conformal mapping allows us to build an
equivalent problem with a much easier boundary condition, the new
interface being plane instead of rough.
Let us associate to the  stream-function
$\Psi$ in the actual space ${\cal E}$ the real potential $\Phi$ in
the half-plane ${\cal D}$:
\begin{equation}
\Phi(z)=\Psi \circ \Omega^{-1}(w) \;,
\end{equation}
where $\Omega$ maps ${\cal D}$ onto ${\cal E}$.  We can thus define
the following equivalent problem in the new geometry:
\begin{equation} \label{DE}
\begin{array}{ll}
&\left\{
\begin{array}{ll}
\nabla_w^4 \Psi (w)=0 &\ \\
{\bf \nabla}_w \Psi (w)=0 &{\rm on  \ } \partial {\cal E} \\
\Psi(w) \sim \frac{\displaystyle 1}{\displaystyle 2} v^2 &{\rm as \ }
v \to - \infty
\end{array}
\right. \vspace{5pt}
\\
\Leftrightarrow
&\left\{
\begin{array}{ll}
\nabla_z^2 \left[ \frac{\displaystyle \nabla_z^2 \Psi (z)}{\displaystyle \vert
    \Omega ' (z) \vert ^2}\right] =0 &\ \\
{\bf \nabla}_z \Phi (z)=0 &{\rm on  \ } \partial {\cal D} \\
\Phi(z) \sim \frac{\displaystyle 1}{\displaystyle 2} y^2 &{\rm as  \ }
y \to - \infty
\end{array}
\right.
\end{array}
\end{equation}
In the case of a simple harmonic equation, building the conformal
mapping gives immediatly the complete solution; this is unfortunately
no longer true in the case of a
bi-harmonic equation. One can see
that the original equation is changed into a linear equation with
non-constant coefficients. The latter equation is directly related to
the mapping function $\Omega$. We show in the following that this
difficulty can be circumverted.  Let us recall that in addition to the
above described conditions, the new potential $\Phi$ has to be real
and $2 \pi$-periodic in $x$. Besides, the boundary condition can be
made simpler taking into account that the interface is now plane.
$\Phi$ being defined apart from an additive constant, we can write
that it obeys:
\begin{equation}
\begin{array}{lcl}
{\bf \nabla}_z \Phi = {\bf 0} \quad {\rm on \ } \partial {\cal D} 
&\Leftrightarrow 
&\left\{ 
\begin{array}{ll}
\Phi &= 0 \quad {\rm on \ }  \partial {\cal D}\\
\frac{\D \partial \Phi}{\D \partial y} &=0 \quad {\rm on \ } 
\partial {\cal D} \quad .
\end{array} \right.
\end{array} 
\end{equation}
The bi-harmonic potential $\Phi$ can always be written in terms of two
holomorphic functions $F$ and $H$ such that:
\begin{equation}
\Phi (z) = \Omega (z) {\overline F(z)} +F (z) {\overline \Omega (z)}
+H(z) + {\overline H(z)} \;.
\end{equation}
In the following we split both functions into a purely periodic part
(denoted by the index $p$) and a non-periodic part. Taking into
account the desired behaviors at infinity provides:
\begin{equation}
\begin{array}{l}
F(z) ={\D \frac {1}{8} z + F_p(z)} \;, \\
H(z) ={\D -\frac{1}{8} z^2 + H_1(z) + H_p(z)}\;, 
\end{array}
\end{equation}
 with
\begin{equation}
\begin{array}{ll}
{\D  F_p (z)= \sum_{n\ge 0} f_n e^{-inz}},
&{\D  H_p (z)= \sum_{n\ge 0} h_n e^{-inz} }\;, 
\end{array}
\end{equation}
and $H_1$ is $z$ times a $x$-periodic function.
The lateral periodicity of $\Phi$ forbids the occurrence of terms
proportional to polynoms of the real variable $x$, hence:
\begin{equation}
H_1(z)=-z \left( F_p(z) + {1 \over 8} \omega (z) \right)\;, 
\end{equation}
and $\Phi$ becomes then:
\begin{equation} \label{phi(z)}
\begin{array}{ll} \vspace{4pt}
  \Phi (z)= &{\D \frac{1}{2} y^2 + \sum_n h_n e^{-inz} +\sum_n
  {\overline h_n} e^{in{\overline z}} }
\\ \vspace{4pt}
&{\D +2iy \sum_n\left({\overline f_n} +\frac{1}{8}\overline \omega_n
  \right) e^{in{\overline z}} } 
\\ \vspace{4pt}
&{\D -2iy\sum_n \left(f_n
+\frac{1}{8}\omega_n \right) e^{-inz} } 
\\ \vspace{4pt}
&{\D +\left.\sum_{n\ge 1}
    \sum_p \left( f_p {\overline \omega_{n+p}} +\omega_{p} {\overline
        f_{n+p}} \right) e^{inx} e^{\left(n+2p \right)y}\right. }
\\   \vspace{4pt}
&{\D +\sum_{n\ge 1} \sum_p \left( {\overline \omega_p}
    f_{n+p} +{\overline f_{p}} \omega_{n+p} \right) e^{-inx}
  e^{\left(n+2p \right)y} }
\\ 
&{\D +\sum_n \left( \omega_n
    {\overline f_n} + {\overline \omega_n } f_n \right) e^{2ny}}\;.
\end{array} 
\end{equation}
We can deduce from the latter expression the partial derivative of
$\Phi$ with  $y$:
\begin{equation}
\begin{array}{ll} \vspace{4pt}
  \frac{\D \partial \Phi}{\D \partial y} (z)&= y + \sum_n n h_n
  e^{-inz} +\sum_n n {\overline h_n} e^{in{\overline z}} \\ 
  \vspace{4pt}&\hspace{-16pt}+2iy \sum_n n \left({\overline f_n} +\frac{\D 1}{\D
      8}{\overline \omega_n} \right) e^{in{\overline z}} -2iy \sum_n n
  \left(f_n +\frac{\D 1}{\D 8}\omega_n \right) e^{-inz} \\ 
  \vspace{4pt}&\hspace{-16pt}+2i \sum_n \left({\overline f_n} +\frac{\D 1}{\D
      8}{\overline \omega_n} \right) e^{in{\overline z}} -2i \sum_n
  \left(f_n +\frac{\D 1}{\D 8}\omega_n \right) e^{-inz} \\ 
  \vspace{4pt}&\hspace{-16pt}+\left.\sum_{n \ge 1} \sum_p \left( n+2p \right) \left( f_p
      {\overline \omega_{n+p}} +\omega_{p} {\overline f_{n+p}} \right)
    e^{inx} e^{\left(n+2p \right)y}\right. \\ 
\vspace{4pt}&\hspace{-16pt}+\sum_{n \ge 1}
  \sum_p \left( n+2p \right) \left( {\overline \omega_p} f_{n+p}
    +{\overline f_{p}} \omega_{n+p} \right) e^{-inx} e^{\left(n+2p
    \right)y} \\ 
\vspace{4pt}&\hspace{-16pt}+\sum_n 2n \left( \omega_n {\overline
      f_n} + {\overline \omega_n } f_n \right) e^{2ny} \;.
\end{array}
\end{equation}

The holomorphic functions $\Omega_p$, $F_p$ and $H_p$ being
$2\pi$-periodical, they can be developed using the basis of the
functions $\{e^{-ikz}\}$. Besides, the interface $\partial {\cal D}$
being simply the $x$-axis, the restriction of these holomorphic
functions to the boundary can be written in $\{e^{-ikx}\}$. The
boundary condition problem can then be written by cancelling out the
projections of $\Phi$ and $\partial_y \Phi$ on the functions-vectors
$\{e^{-ikx}\}$. Keeping only the first $N$ modes, we obtain $2N$
equations which allow to obtain the $2N$ components $f_n$ and $h_n$.
The projection of the boundary condition on the function-vector $1$
($k=0$) gives first:
\begin{equation} \label{f0}
\left \{
\begin{array}{ll}\vspace{4pt} 
\Phi : &{\D \sum_n \left( \omega_n
        {\overline f_n} + {\overline \omega_n } f_n \right) +h_0
      +{\overline h}_0=0} \;, \\ 
\partial_y \Phi : &{\D \sum_n 2n \left(
        \omega_n {\overline f_n} + {\overline \omega_n } f_n \right)
      +2i\left({\overline f_0 }+\frac{\D 1}{\D 8}{\overline
          \omega_0}\right)} \\
&{\D -2i\left( f_0 +\frac{1}{8} \omega_0
      \right)=0 }\;.
\end{array}\right.
\end{equation}
We have then for the function-vectors $\{e^{-ikx}\}$ with  $k \ne 0$:
\begin{equation}
\left \{ 
\begin{array}{ll}\vspace{4pt}
\Phi : & {\D \sum_p \left({\overline
\omega_p} f_{k+p} +{\overline f_{p}} \omega_{k+p}\right)+h_k =0}  \;,
  \\
\partial_y \Phi : 
&{\D \sum_p \left( k+2p
\right) \left( {\overline \omega_p} f_{k+p} +{\overline f_{p}} \omega_{k+p}
\right) +k h_k }
\\
 &{\D -2i \left(f_k +\frac{1}{8}\omega_k \right) =0 } \;.
\end{array}\right.
\end{equation}
We can thus write the following linear system:
\begin{equation}
\left \{ \begin{array}{l}\vspace{4pt}
{\D f_k + \frac{\D 1}{\D 8}\omega_k  +i  \sum_p p\left( {\overline
\omega_p} f_{k+p} +{\overline f_{p}} \omega_{k+p} \right) =0}  \;, \\
{\D h_k =- \sum_p \left({\overline \omega_p} f_{k+p} -{\overline f_{p}}
\omega_{k+p} \right)} \;.
\end{array} \right.
\end{equation}

\begin{figure}
  \centerline{
    \epsfig{file=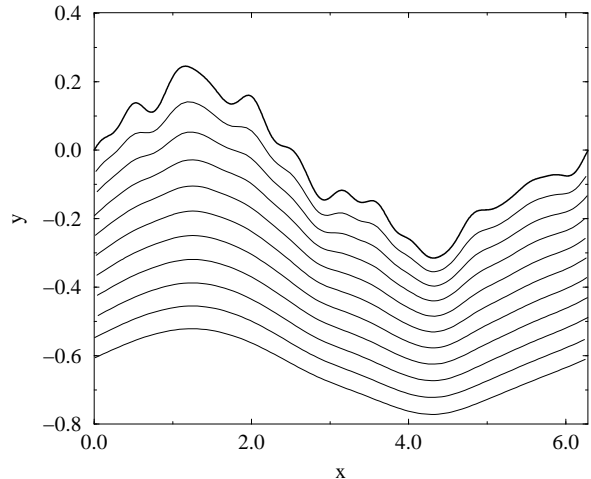,width=0.85\hsize,angle=-90}}
    \caption{Stream lines of a Stokes flow along a rough
      surface.} 
\label{Stokes1}
\end{figure}

\begin{figure}
  \centerline{
    \epsfig{file=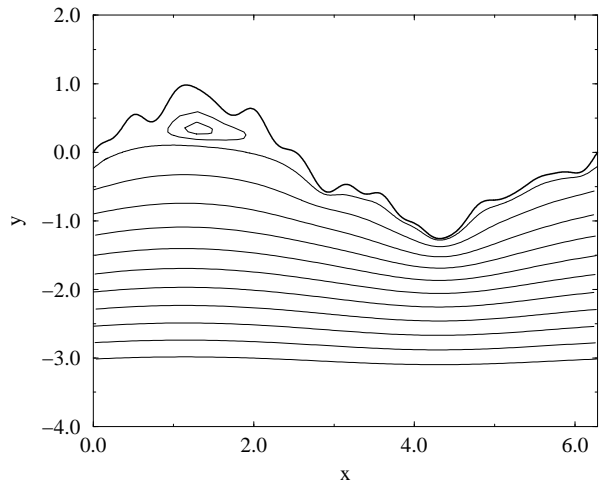,width=0.85\hsize,angle=-90}}
    \caption{Stream lines of a Stokes flow along the same rough
      surface as above but four times rougher. We observe that a
      recirculation flow appears in the deepest cavity.} 
\label{Stokes2}
\end{figure}

The coefficients $\{f_k\}$ are solutions of the first equation, a $N
\times N$ linear system. The coefficients $\{h_k\}$ are easily deduced
from the $\{f_k\}$. Once the conformal mapping is known, the numerical
resolution of the Stokes equation just requires the inversion of the
linear system. Let us note that the latter system is easy to invert,
which would not have been the case if we have written the boundary
condition in the $w$-space. The latter method was recently used by
Tuck et Kouzoubov \cite{Tuck95}. In the cited reference, they use
expansions in a basis of $\{e^{-ky}cos(kx)\}$ and write a linear
system discretizing the rough interface in $N$ points. If this method
is efficient in the small slope limit, it becomes however
unpracticable for large slopes. The procedure requires thus the
numerical inversion of a matrix consisting of terms of order $e^{\pm
  NA}$ where $A$ is the roughness amplitude and this becomes
practically difficult or unprecise as soon as the product $NA$
increases. The conformal mapping avoids such numerical difficulties
since the boundary is plane in the equivalent domain.  One can see on
figures \ref{Stokes1} and \ref{Stokes2} maps of the stream function
for a stationary Stokes flow for two rough surfaces identical up to a
dilation of factor $4$ in the vertical direction.  The stream lines
closely follow the smooth interface (Fig. \ref{Stokes1}) while an eddy
appears in the largest depression of the roughest interface (Fig.
\ref{Stokes2}).

\subsection{Equivalent plane boundary}

We now introduce the notion of an equivalent plane no-slip boundary in the
framework of a stationary Stokes flow. Our aim is here to replace the
no-slip condition on the rough boundary by a no slip condition on an
equivalent plane boundary, the stream function remaining unchanged at
infinity. Let us recall that like in the case of a rough electrode
which was studied in Ref. \cite{VdbR96a}, nothing prescribes the plane
equivalent boundary to lie at the average height of the rough one. In
harmonic problems, the dissymetry between peaks and cavities due to
point effects leads the equivalent plane boundary to be shifted
towards the peaks. We shall see in the following that the same
conclusion holds in the case of the stationary Stokes flow. An
illustration of this disymmetry naturally emerges with the occurrence
of little eddies in pronounced depressions of the rough boundary. We
have chosen in this text to develop the paradigm of no-slip plane
equivalent boundary but it is naturally also possible to consider a
plane boundary fixed at the average plane with a roughness dependent
slip boundary. The conclusion previously mentioned about the place of
the equivalent boundary (nearer from the peaks than from the cavities)
is then expressed by a reversal flow condition.  We show in the
following that the conformal mapping method gives a natural way to
compute the vertical shift of the plane equivalent boundary. We
compare these results with a perturbative solution that can be
directly computed from the Fourier coefficients of the interface.

\subsubsection{Conformal mapping approach}

As already mentioned, the stream function $\Psi(u,v)$ of the Stokes
flow is entirely determined by the following conditions: 
\begin{equation}  \label{eqperbi}
\left\{ 
\begin{array}{rlcc}
\nabla_w^4 \Psi (w)& = 0\quad & {\rm in}\quad & {\cal E} \\ 
{\bf \nabla}_w \Psi (w) & ={\bf 0} \quad & {\rm on}\quad & \partial{\cal E} \\ 
\Psi(w) & \sim  \frac{1}{2}v^2 \quad & {\rm as}\quad & v \to - \infty
\end{array}
\right.
\end{equation}

If we now return to the solution obtained via conformal mapping, we have:

\begin{equation} \label{stream}
\begin{array}{ll} \vspace{4pt}
  \Phi (z)= &{\D 1 \over \D 2} y^2 + \sum_n h_n e^{-inz} +\sum_n
  {\overline h_n} e^{in{\overline z}} \\ \vspace{4pt}&+2iy
  \sum_n\left({\overline f_n} +\frac{\D 1}{\D 8}\overline \omega_n
  \right) e^{in{\overline z}} \\ 
\vspace{4pt}&-2iy\sum_n \left(f_n +\frac{\D 1}{\D
      8}\omega_n \right) e^{-inz} \\ \vspace{4pt}&+\left.\sum_{n\ge 1}
    \sum_p \left( f_p {\overline \omega_{n+p}} +\omega_{p} {\overline
        f_{n+p}} \right) e^{inx} e^{\left(n+2p \right)y}\right. \\ 
  \vspace{4pt}&+\sum_{n\ge 1} \sum_p \left( {\overline \omega_p}
    f_{n+p} +{\overline f_{p}} \omega_{n+p} \right) e^{-inx}
  e^{\left(n+2p \right)y} \\ \vspace{4pt}&+\sum_n \left( \omega_n
    {\overline f_n} + {\overline \omega_n } f_n \right) e^{2ny}\;.
\end{array}
\end{equation}

By construction, $\Phi(z)$ is such that $\Psi(w)$ is bi-harmonic in
${\cal E}$ and fulfills the no-slip boundary condition at the rough
interface. Let us now build the stream function $\Psi_{eq}$ associated
with a plane interface located at $v=H$; we have immediatly:
\begin{equation}
\Psi_{eq}(w)=\frac{1}{2}\left( v-H \right)^2 \;,
\end{equation}
and its associated function in the half-plane ${\cal D}$ is
\begin{equation}
\Phi_{eq}(z)=\frac{1}{2}\left[\mathop{\Im m}\Omega(z)-H \right]^2 \;,
\end{equation}

that becomes at infinity:
\begin{equation}
\begin{array}{lr}
{\D   \Phi(z)= \frac{ 1}{2}y^2 +
  \frac{1}{2} y \mathop{\Im m} \left( \omega_0 +8f_0\right) 
    +{\cal O}(1)} &{\rm as \ 
    } y \to -\infty \;, \\
{\D   \Phi_{eq}(z)= \frac{1}{2}y^2 + y \left[\mathop{\Im m}(\omega_0)-H \right]+{\cal O}(1)} &{\rm as \ 
    } y \to -\infty \;.
\end{array}
\end{equation}
Taking into account Eq.(\ref{f0}) which specifies the expression
of  $f_0$, the identity between $\Phi$ and $\Phi_{eq}$ defines the
value of $H$
\begin{equation}
  H=\mathop{\Im m}(\omega_0) +2\sum_{n} n \left( \omega_n {\overline f_n} + {\overline \omega_n}
    f_n \right) \;.
\end{equation}

\subsubsection{Perturbative approach}

Let us consider an interface of amplitude, say $A$, with
characteristic length $\lambda$ such that the profile is statistically
symmetrical. When we seek for the location of the plane equivalent
electrode, we expect two different behaviors  i) a
dependence on $A^2/\lambda$  in case of small amplitude or low spatial
frequency ii) a linear dependence on
$A$ in case of large amplitude or high spatial frequency. The latter
behavior comes directly from the fact that the equivalent plane reaches at maximum the level
of the highest peaks.
In case of small slopes, it is easy to show that the correction from the
average plane is of order $A^2/\lambda$. The deviation $H$ is
naturally normalized by the amplitude of roughness $A$. The ratio
$H/A$ has then to be a function of the two only characteristic lengths
of the system, $A$ and $\lambda$ and can be expanded in the limit of
small slopes:
\begin{equation}
\frac{H}{A}=\phi( \frac{A}{\lambda})=a_0 +a_1
\frac{A}{\lambda} +a_2 \left(\frac{A}{\lambda}\right)^2
+ {\cal O}\left(\frac{A}{\lambda}\right)^3 \;.
\end{equation}
A simple symmetry about the mean plane has to leave $H$ unchanged,
since this symmetry is equivalent to a transformation of $A$ into
$-A$, $a_0$ and $a_2$ have to be zero and:
\begin{equation}
H=a_1\frac{A^2}{\lambda} + {\cal
O}\left(\frac{A^4}{\lambda^3}\right)
\end{equation}

A detailed perturbative analysis can be built in the case of a simple
sine interface \cite{Tuck95}. Writing a perturbative solution in the
conformal mapping formalism allows us to deal with any rough
interface. Following (\ref{phi(z)}) we have:
\begin{equation}
\begin{array}{ll}
\Phi(z)=&{\D \frac{1}{2}y^2 +4y {\Im m}\left(F_p(z)+\frac{1}{8}\omega(z)
\right) }
\\ \vspace{4pt} 
&{\D +2 {\Re e}\left(\omega(z)\overline{F_p(z)} + H_p(z)\right)} \;.
\end{array}
\end{equation}
By construction, $\omega$, $F_p$ and $H_p$ are of  order $A$, $A$ being the roughness amplitude. At first order the no-slip boundary
condition becomes:
\begin{equation}
\begin{array}{lcl}\vspace{4pt}
\Phi(x)=0 &\iff &{\Re e}\left(H_p^{(1)}(x)\right)=0 \;,\\
{\D \frac{\partial \Phi}{\partial y} (x)=0} &\iff &4 {\D {\Im
  m}\left(F_p^{(1)}(x)+\frac{1}{8}\omega^{(1)}(x) \right)}\\ 
& &{\D +2{\Re
  e}\left(\frac{\partial H_p^{(1)}}{\partial y}(x)\right)=0 }\;.
\end{array}
\end{equation}
The holomorphic functions being bounded, the resolution of these
Hilbert problems gives immediatly:
\begin{equation}
H_p^{(1)}(z)=0\;, \quad F_p^{(1)}(z)=-\frac{1}{8} \omega^{(1)}(z) \;.
\end{equation}
Using the iterative algorithm briefly presented in section II (see
Appendix for details), a first order approximation of the coefficients
$\omega_k$ is
\begin{equation}
\omega_k=i \frac{\tilde{h}_k}{N} + {\cal O}(A^2) \;,
\end{equation}
where $\{\tilde{h}\}$ is the $2N$ discrete Fourier transform of the
real function $h$ associated with the rough interface. Using the
expression of ${\Im m}(\omega_0)$ derived in section V-B of
Ref. \cite{VdbR96a}, we can write:

\begin{equation}
H=-\frac{1}{N^2}\sum_{k>0}k\left|\tilde{h}_k \right|^2 \;.
\end{equation}

This result is consistent with the one proposed in Ref. \cite{Tuck95}
for a backflow slip condition on the mean plane of the interface.
In the particular case of a pure sine profile of amplitude $A$ and
wavelength $\lambda$, we recover:
\begin{equation}
H=-2\pi \frac{A^2}{\lambda}\;.
\end{equation}

\begin{figure}
  \centerline{
    \epsfig{file=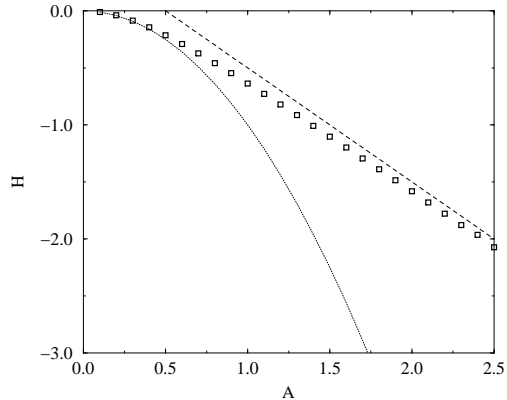, height=0.85\hsize,angle=-90}}
    \caption{Shift from the mean plane of the plane equivalent
      boundary for a Stokes flow in case of sine interface of varying
      amplitude $A$. The dotted line corresponds to the second order
      perturbative expansion and the symbols to computations using
      conformal mapping. The dashed line has a slope -1.}
\label{Stokesequivsin}
\end{figure}

\begin{figure}
  \centerline{
    \epsfig{file=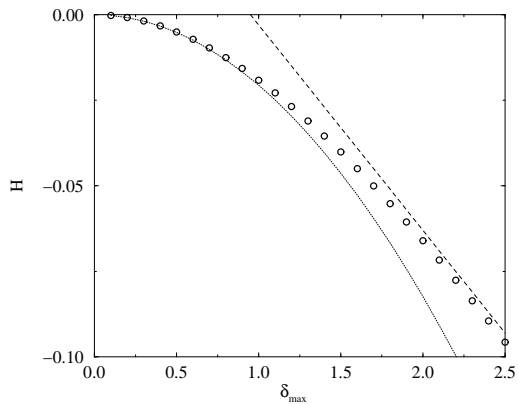, height=0.85\hsize,angle=-90}}
    \caption{Shift from the mean plane of the plane equivalent
      boundary for a Stokes flow in case of self-affine interface of varying
      amplitude. We use in abscissa the maximum local slope
      $\delta_{max}$. The dotted line correspond to the second order
      perturbative expansion and the symbols to computations using
      conformal mapping. The dashed line has a slope -0.06. The
      surface has been built with 64 Fourier modes and we used 256
      modes in the solution based on conformal mapping.}
\label{Stokesequiv}
\end{figure}

It has to be noted these first order results are exactly identical to
those obtained in the case of a rough electrode (up to a factor 2)
despite the fact we had here to solve a biharmonic equation instead of
a simple harmonic one.  On figures \ref{Stokesequivsin} and
\ref{Stokesequiv} we have plotted both results of the perturbative
solution and of the conformal mapping calculation in case of a sine
interface and of a self affine interface of roughness exponent 0.8
built with 64 Fourier modes. We check that the perturbative
calculations correctly fit the results for small slopes (up to 0.5).
For larger slopes, the perturbative expression overestimates the
deviation whose behavior becomes progressively linear. Note that our
numerical method allowed us to reach local slope values up to 2.5.
This maximum slope can be increased by using more Fourier modes in the
mapping function (we used 256 modes in the present calculation).

\section{Elasticity}
In this section, we analyse the effect of a slight
surface roughness on the stress distribution in an elastic medium. We
emphasize here the case of a semi-infinite material submitted to a
uni-axial tension. Although very elementary, this simple
model illustrates an effect which has been suggested to be responsible
for the mechanical strength of glass fibers.
Recent experimental results \cite{Guilloteau} suggest that the
nanometric roughness at the surface of glass fibers of diameter a few
micrometers could be responsible for the decrease of the tensile
resistance by a factor of about 5. In uniaxial tension, the resistance
of a fiber is directly related to the distribution of maximum positive
principal stresses. Using a simple perturbative expansion, we show
that in the limit of small slopes, the surface stress can be directly
computed from the Hilbert transform of the local slope. 

The Weibull law \cite{Weibull39} usually gives a
correct description of failure statistics in a wide range of brittle
materials. This phenomenological law is partly based on the
identification of the weakest link in the material (whose size is
supposed to be that of the smallest defects). We will see in the
following that for a self-affine surface (such as measured for glass
fibers by AFM techniques \cite{Guilloteau}) the statistical
distribution of tensile stresses at the boundary displays a power law
behavior, which naturally implies the validity of Weibull statistical
failure distribution. Moreover the exponent of this power law, {\it
  i.e.} the Weibull modulus, continuously depends on the roughness
amplitude.

\subsection{General solution}
In plane stress or plane strain conditions, the stress tensor $[\sigma]$ can be completely
represented by a unique real function, named the Airy function:
\begin{equation}
[\sigma] = \left[\matrix{\sigma_{uu} &\sigma_{uv} \cr \sigma_{uv}
&\sigma_{vv}} \right] 
\quad 
{\rm where} \quad
\left\{ \begin{array}{lc}
\vspace{2pt}
\sigma_{uu}=&\frac{\D \partial^2 \Psi}{\D \partial v^2}\\\vspace{2pt}
\sigma_{uv}=&-\frac{\D \partial^2 \Psi}{\D \partial u\partial v}\\
\sigma_{vv}=&\frac{\D \partial^2 \Psi}{\D \partial u^2} 
\end{array} \right. 
\end{equation}
This form directly comes from the stress balance in absence of
external forces {\it i.e.} ${\rm div} [\sigma]=0$. In the framework of 2D
elasticity in an isotropic medium, the Airy function obeys:
\begin{equation}
\Delta\Delta \Psi = 0 \quad {\rm in} \enskip {\cal E} \;.
\end{equation}
The stress tensor being computed from two successive derivations of
the bi-harmonic function $\Psi$, the latter is only defined apart
from a linear function in $u$ and $v$.  In the following, we consider
free 
boundary conditions $[\sigma]{\bf n}={\bf 0}$
 and we impose a uni-axial tension at infinity. Let ${\bf n}$
denote a unit vector normal to the interface, we have:
\begin{equation}
\left\{
\begin{array}{llll}
\vspace{4pt}
[\sigma ] {\bf n}  &= &{\bf 0} &{\rm on \ } \partial {\cal E}\;,\\
 \left[ \sigma  \right]  &\rightarrow 
    &\left[
    \begin{array}{cc}
     1&0\\
     0&0
    \end{array}
    \right]
&{\rm as \ } y
\rightarrow -\infty \;.
\end{array}\right.\end{equation}
The Airy function $\Psi$ is thus such that:
\begin{equation}
\left\{\begin{array}{clll}
\vspace{4pt}
n_u \frac{\D \partial^2 \Psi}{\D \partial v^2} -n_v \frac{\D \partial^2
\Psi}{\D \partial u\partial v}&= &0 &{\rm on \ } \partial {\cal E}\;,\\
\vspace{4pt}
n_u \frac{\D \partial^2\Psi}{\D \partial u\partial v}-n_v \frac{\D \partial^2
\Psi}{\D \partial u^2} &= &0 &{\rm on \ } \partial {\cal E}\;, \\
\Psi &\sim &\frac{\D 1}{\D 2} v^{2}  &{\rm as\ } v \rightarrow -\infty \;.
\end{array}\right.
\end{equation}
At any point of the interface $\partial {\cal E}$, it is possible to
give a parametric representation of the tangential and normal vectors ${\bf t}$
 and  ${\bf n}$:
\begin{equation}
{\bf t}\left(w\right) = {{\Omega ' \left(x\right)}\over
{\left|\Omega ' 
\left(x\right)\right|}}
\enskip ,\quad 
{\bf n}\left(w\right) = {{i \Omega '\left(x\right)}\over {\left|\Omega '
\left(x\right)\right|}}\;.
\end{equation}
The boundary conditions at the interface can be rewritten:
\begin{equation}
\begin{array}{rll}\vspace{4pt}
&[\sigma] {\bf n} =\enskip 0 \quad &{\rm on}\enskip \partial {\cal
  E}\\ \vspace{4pt} 
\Leftrightarrow &\left({\bf t} \cdot \nabla_w \right)\nabla_w
\Psi =\enskip 0 \quad &{\rm on}\enskip \partial {\cal E} \\ \vspace{4pt}
\Leftrightarrow & \nabla_w \Psi = Cst. \enskip=\enskip 0 &{\rm
  on}\enskip \partial {\cal E}\;.
\end{array}
\end{equation}
We can choose the constant to be zero since the Airy function is only
defined apart from an affine function. This leads to:
\begin{equation}
\begin{array}{rll}
\vspace{4pt}
&\frac{\D \nabla_z \Phi}{\D \overline{\Omega '}\left(z\right)} =0 & 
\quad {\rm on}\enskip \partial D \\ 
\Leftrightarrow &{\nabla_z \Phi}=0 & \quad{\rm on}\enskip
\partial D \;.
\end{array} 
\end{equation}
It turns out that the boundary condition at the interface is exactly
the same as the one we have encountered for the no-slip condition in a
Stokes flow. The Airy function is thus identical to the above derived stream
function for Stokes flow.

\subsection{Surface stress distribution}

\subsubsection{Perturbative approach}
\begin{figure}
  \centerline{
    \epsfig{file=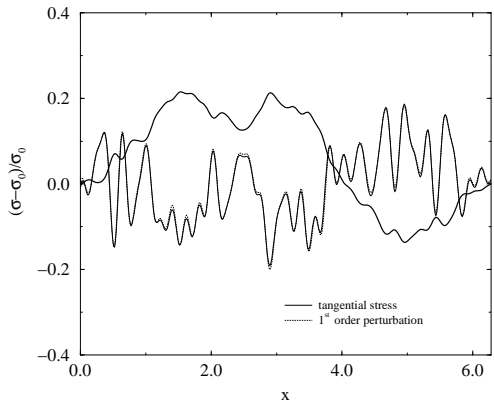, height=0.85\hsize,angle=-90}}
    \caption{Stress and height profiles on the rough interface of a 2D
      medium submitted to uni-axial tension. The bold line represents
      the height profile (of maximum slope 0.1) dilated by a factor 5.
      The solid line gives the stress distribution obtained by the
      complete conformal mapping computation, the dotted line is a
      first order expression of the stress which is directly obtained
      from the Hilbert transform of the local slope of the interface.}
\label{cont01}
\end{figure}

\begin{figure}
   \centerline{
    \epsfig{file=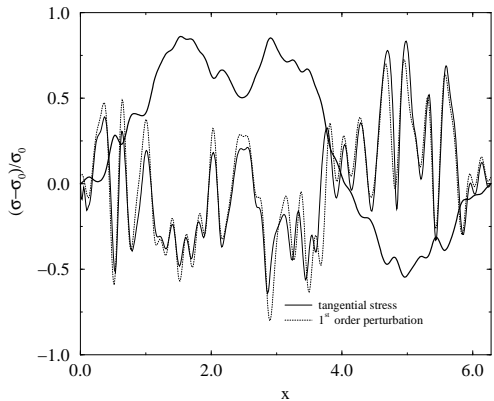, height=0.85\hsize,angle=-90}}
    \caption{Same figure as above with an interface four times
      rougher. One observes that in large curvature areas, the first order
      approximation does no longer suffice to represent precisely the
      local stress. The height profile (of maximum slope 0.4) dilated by a factor 5.}
\label{cont04}
\end{figure}

The normal stress being zero at the interface, the first order
expression derived in the previous section gives us 
the following result for the principal (tangential) stress at the interface:

\begin{equation}
\begin{array}{ll}
  \vspace{6pt} 
\sigma_{tt}&={\D \Delta \Psi(w)=\frac{ \Delta
      \Phi(x)}{\vert{\Omega'}(x)\vert^2} } \\ 
  \vspace{6pt} 
&{\D  = 1 -2 {\Re e} \left[ \omega^{(1)'} (x) \right]}\\
  \vspace{6pt} 
&{\D  = 1 -2 \sum_{k=-n+1}^{k=n}i \frac{k}{|k|} (-ik) \frac{\widetilde{h_k}}{2n}
  e^{-ikx}} \\
&{\D  = 1 -2 {\cal H}[\varphi'](x)} \;,
\end{array}
\end{equation}
where ${\cal H}[...]$ stands for the Hilbert transform oprerator on the
real axis. At first order, we find  that
the tangential surface stress deviation from its mean value is
proportional to 
{\it  the Hilbert
transform of the local slope}. Fig. \ref{cont01} and \ref{cont04} give
two examples of stress distribution on surfaces of repectively maximum
slope 0.1 and 0.4. In both cases, we can see that the stress
fluctuations are much greater than the height fluctuations (which have
been dilated by a factor 5 in the figures). Note that (taking into
account the dilation of the height profile) the stress fluctuations
are very large compared with those of the height. In the context of
rupture, the relevant parameter being the maximum stress, one can
understand that a very modest roughness can be responsible for a
dramatic decrease of the material resistance.

We observe a good agreeement between the stress profile computed by
conformal mapping and the perturbative result (related to the Hilbert
transform of the slope) for the smoothest interface but this is no
longer the case for the roughest one especially in the area where the
curvature is large. Higher order perturbative terms are thus necessary
to recover the actual stress.

\subsubsection{Statistical results}

\begin{figure}
   \centerline{
    \epsfig{file=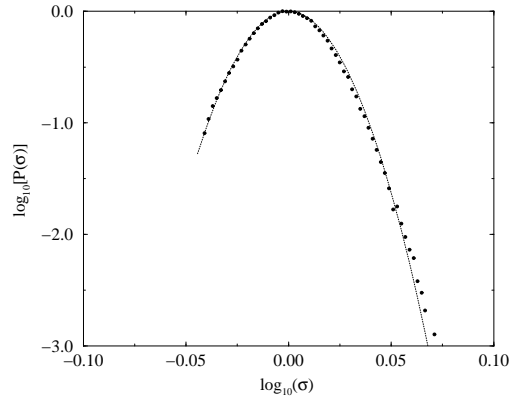, height=0.85\hsize,angle=-90}}
    \caption{Surface stress distribution for self-affine surfaces of
      roughness exponent $\zeta=0.8$ and of roughness amplitudes
      $A=0.05\epsilon_{max}$, $\epsilon_{max}$ is the amplitude such that
      the maximum local slope is equal to $1$. The self-affine
      interfaces have been built with 32 Fourier modes and the results
      averaged on 1500 surfaces. We can check that the distribution
      (symbols) is well fitted by a parabola, which shows that the
      stress is log-normally distributed. }
\label{lognormal}
\end{figure}

Let us now turn to the study of stress distributions on self-affine
surfaces. The latter are designed to present a Gaussian height
distribution {\it i.e.} their Fourier coefficients are
$\widetilde{h_k}=A \epsilon_k k^{-\frac{1}{2}-\zeta}$ where
$\epsilon_k$ is a Gaussian random variable of zero mean and unit
standard deviation. In case of small slopes, the validity of the first
order perturbative result suggests thus that the stress distribution
is Gaussian. On Fig. \ref{lognormal} we have plotted in log-log scale
the surface stress distribution obtained for 1500 self-affine surfaces
of roughness exponent $\zeta=0.8$ and of maximum slope 0.05, one can
check that, as expected, the distribution is well fitted by a
parabola.

\begin{figure}
   \centerline{
    \epsfig{file=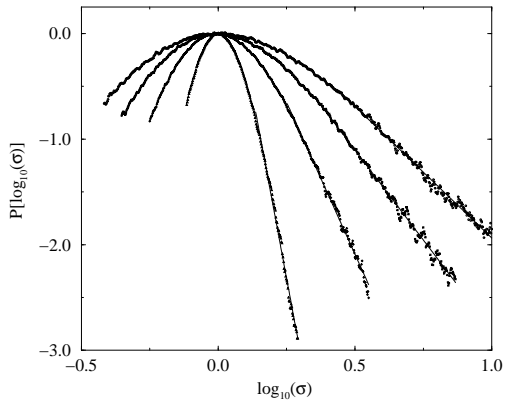, height=0.85\hsize,angle=-90}}
    \caption{Surface stress distribution for self-affine surfaces of
      roughness exponent $\zeta=0.8$ and of roughness amplitudes
      $A=0.2\epsilon_{max}$ ($\triangle$), $0.4\epsilon_{max}$
      ($\diamond$), $0.6\epsilon_{max}$ ($\sqcap$) and
      $0.8\epsilon_{max}$ ($\circ$), $\epsilon_{max}$ is the amplitude
      such that the maximum local slope is equal to $1$. The
      self-affine interfaces have been built with 64 Fourier modes and
      the results averaged on 1000 surfaces. For each distribution, a
      bold line shows the power law behavior obtained for large
      stresses.}
\label{powerlaw}
\end{figure}

\begin{figure}
  \centerline{
    \epsfig{file=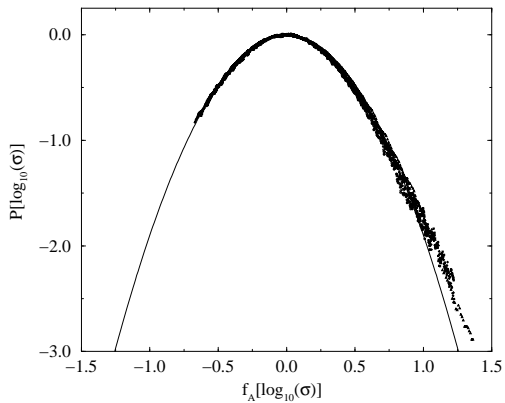, height=0.85\hsize,angle=-90}}
    \caption{Same distributions as in the previous figure in the reduced variable
      $f_A(g)=\left(\sqrt{1+2Kg}-1 \right)/KA$
      where $g=\log_{10}(\sigma)$ and $K=0.25$. The data collapse onto a simple
      parabola, which shows the Gaussian character of the
      distribution of  $f_A(g)$}
\label{collapse}
\end{figure}

In case of larger slopes we have seen that the agreement between the
first order perturbative results and the complete computation becomes
poorer. We can then wonder if the log-normal behavior of the
distribution is preserved.  On Fig. \ref{powerlaw}, we have plotted
the surface stress distribution for four self-affine surfaces of
roughness exponent $\zeta=0.8$ and of respective maximum slopes 0.2,
0.4, 0.6 and 0.8. These results were obtained by averaging the data
obtained from 1000 different surfaces each defined with 64 Fourier
modes. We can see a clear power law like behaviour for large stress
amplitudes. The slopes we can measure are very dependent on the
roughness amplitude.  The interpretation of these new numerical
results requires a perturbative analysis that we have not developed
yet for this bi-harmonic problem.  We have however performed a similar
analysis  in the case of a harmonic field on self-affine
interfaces \cite{VdbR96distri}. It turned out that in a similar
fashion as the one presented above, the field distribution law present
a power law tail with an exponent $\tau \propto A^{-2}\ell^{1-\zeta}$
where $A$ is the roughness amplitude, $\ell$ the spatial lower cut-off
of the self-affine domain and $\zeta$ the roughness exponent. Calling
$g$ the logarithm of the field, the latter result was derived showing
that the reduced variable
\begin{equation}
f_A(g)=\left(\sqrt{1+2Kg}-1 \right)/KA
\end{equation}
follows a Gaussian distribution ($K \simeq 2$ for harmonic problems).
Using $K=0.25$, we can check indeed (see Fig. \ref{collapse}) that all
data obtained from our calculations collapse on a same parabola in
log-log scale. These numerical results indicate that a same scaling
applies for both harmonic and bi-harmonic fields on self-affine
surfaces. A detailed analysis of these statistical properties will be
addressed elsewhere.

\section{Conclusion}
After introducing a conformal mapping technique that allowed for a
detailed study of harmonic field in the vicinity of rough boundaries
\cite{VdbR96a}, we have extended in this paper the use of this method
to the study of bi-harmonic fields. We have given a general solution
to problems such that the Stokes flow over a rough surface and the
stress distribution in a medium (bounded by a rough interface) in
uni-axial tension. Besides the knowledge of the mapping function
(obtained using a simple iterative algorithm), this solution only
requires the linear inversion of a well conditioned matrix. The
determination of the mapping function being only limited by the
maximum value of the local slope at the interface, the method is well
suited to any kind of single-valued interface.  As an illustration, we
have thus presented results of Stokes flow over self-affine boundaries
whose maximum slope reach 2.5. In the same context of a Stokes flow
over a rough boundary, we pay a special attention to the determination
of the location of an equivalent no-slip plane interface. The
conformal mapping method gives way to a very natural determination of
this quantity. A simple perturbative solution allowed us to retrieve
for it a result proposed by Tuck and Kouzoubov \cite{Tuck95}. In the
context of the plane elasticity, the same perturbative result has
allowed us to show that, in the limit of small slopes, the surface
stress distribution was directly related to the Hilbert transform of
the slope of the interface. This very simple result could be used e.g.
for the evolution of a stress-corrosion front.  The analysis of
statistical results for the principal stress on self-affine surfaces
has shown moreover that the large stress distribution presents a power
law tail whose exponent continuously depends on the roughness
amplitude. Such results could be applied in the context of glass fiber
rupture statistics to provide a fundamental basis for the Weibull law
that is known to describe accurately the rupture statistics.  A
realistic description of these stress distributions require a second
order perturbative analysis which will be presented in a futher study.
We have, however, recently proposed such an approach in case of
harmonic fields \cite{VdbR96distri} where distributions of the same
kind have also been found and justified in two and three dimensions.

\section*{Acknowledgements}
It is a pleasure to acknowledge enlightning discussions with F.
Creuzet, E.J. Hinch and J.R. Willis. We are also grateful to B. Forget
and A. Tanguy for useful help and comments.

\vspace{-5pt}

\section*{Appendix: Iterative algorithm}

In this appendix we present the iterative scheme allowing us to build
the conformal mapping $\Omega(z)=z+\sum_k \omega_k e^{-ikz}$ for any
given single valued rough interface. As exposed in II.B, we are
seeeking for a mapping function $\Omega$ such that $\Omega
\left(\partial{\cal D}\right)=\partial{\cal E}$, where $\partial{\cal
  E}$ is the rough interface and $\partial{\cal D}$ is the x-axis. The
single valued interface $\partial{\cal E}$ being associated to the
real function $h$, this condition is rewritten:
\begin{equation}
\begin{array}{ll}
\vspace{4pt}
&{\D h\left(u \right) = {\rm Im} \left(\sum_{k=0}^\infty \omega_k
e^{-2i\pi kz} \right)}\;,\\
{\rm where} &{\D u=x+{\rm Re} \left(\sum_{k=0}^\infty \omega_k
e^{-2i\pi kz} \right)}\;. \label{eqboom}
\end{array}
\end{equation}

We build the algorithm in the following way: the intermediate
quantities appearing at the $k$th iteration are labelled with a
superscript $(k)$, all functions are decomposed over a set of $2n$
discrete values, the number of Fourier modes is thus limited to $2n$.
We first introduce a series of sampling points $u_j^{(k)}$ with
$j=0,...,\ n-1$ which is initially set to an arithmetic series
$u_j^{(0)}=j\pi /n$. The sampling of $h(u)$ by the $u_j^{(k)}$ gives
the array:
\begin{equation}
h_j^{(k)}=h(u_j^{(k)})  \label{eqalgbeg}
\end{equation}
The discrete Fourier transform of this array is the complex valued array 
\begin{equation}
a_j^{(k)}=\sum_{m=-n+1}^nh_m^{(k)}e^{imj}\;.
\end{equation}
for $-n<j\le n$. The latter is shortly written as: 
\begin{equation}
a^{(k)}={\cal F}[h^{(k)}] \;,
\end{equation}
where ${\cal F}$ denotes the Fourier transform, which will be chosen as the
Fast Fourier Transform (FFT) algorithm, thus imposing that $n$ is an integer
power of 2. The intermediate mapping $\omega ^{(k)}$ is computed from the $%
a^{(k)}$ as: 
\begin{equation}
\left\{ 
\begin{array}{rlcc}
\omega _j^{(k)}= & (i/n)a_j^{(k)} & \qquad {\rm for} & \quad j>0 \\ 
\omega _0^{(k)}= & (i/2n)a_0^{(k)} &  &  \\ 
\omega _j^{(k)}= & 0 & \qquad {\rm for} & \quad j<0 \;.
\end{array}
\right. 
\end{equation}
The latter form is obtained from the identification of Eq.(\ref{eqboom}b)
and the definition of $a^{(k)}$, taking care of the fact that one sum is
over positive index, while the other extends over the interval $[1-n,n]$.
Then, one computes the series:
\begin{equation}
\left\{ 
\begin{array}{rlcc}
b_j^{(k)}= & ia_j^{(k)} & \qquad {\rm for} & \quad j>0 \\ 
b_0^{(k)}= & 0 &  &  \\ 
b_j^{(k)}= & \overline{b_{-j}^{(k)}}=-i\overline{a_{-j}^{(k)}}=-ia_j^{(k)} & 
\qquad {\rm for} & \quad j<0\;.
\end{array}
\right. 
\end{equation}
This linear transformation is shortly noted as: 
\begin{equation}
b^{(k)}={\cal H}[a^{(k)}] \;,
\end{equation}
where ${\cal H}$ is the above detailed transformation. The form of ${\cal G}$
is dictated by Eq.(\ref{eqboom}a) for positive index, and from the fact that
the inverse Fourier transform of $b$ (see below) is real. The new sampling
series is finally obtained from: 
\begin{equation}
u_j^{(k+1)}={\frac{j\pi }n}+{\cal F}^{-1}[b^{(k)}]\;.  \label{eqalgend}
\end{equation}
The equations (\ref{eqalgbeg}-\ref{eqalgend}) define one step in the
algorithm relating $\omega ^{(k+1)}$ to $\omega ^{(k)}$. We note
this step $\omega ^{(k+1)}={\cal T}(\omega ^{(k)})$.
The searched function $\Omega$ is clearly a fixed point of the
transformation ${\cal T}$ defined above in a discretized version. The
uniqueness of the transformation $\Omega$ results from that of the harmonic
field in the domain ${\cal E}$ with an equipotential condition on the
boundary and a constant gradient perpendicular to the boundary at infinite
distance from it. Therefore, the only condition to consider is the stability
of the fixed point. As it can be seen in Ref. \cite{VdbR96a}, this
fixed point is attractive at the only condition that:
\begin{equation}
\max \vert h'(u) \vert <1 \;.
\end{equation}
The procedure described above gives a very fast and efficient way of
building conformal mappings. It has to be noted indeed that this
two-dimensional problem is solved here by the only use of a few
one-dimensional fast Fourier transforms. In cases of slopes locally
overcoming one, it is possible to use
under-relaxation schemes (see details in Ref. \cite{Gut81,Gut83}) or
to decompose the mapping in several substeps such that the
convergence criterium is always fulfilled.


\end{document}